# Towards Mass Adoption of Contact Tracing Apps - Learning from Users' Preferences to Improve App Design

Dana Naous[a], Manus Bonner[a], Mathias Humbert[b] and Christine Legner[a]

[a] Faculty of Business and Economics (HEC), University of Lausanne, Switzerland
[b] armasuisse, Science and Technology, Switzerland

**Abstract.** Contact tracing apps have become one of the main approaches to control and slow down the spread of COVID-19 and ease up lockdown measures. While these apps can be very effective in stopping the transmission chain and saving lives, their adoption remains under the expected critical mass. The public debate about contact tracing apps emphasizes general privacy reservations and is conducted at an expert level, but lacks the user perspective related to actual designs. To address this gap, we explore user preferences for contact tracing apps using market research techniques, and specifically conjoint analysis. Our main contributions are empirical insights into individual and group preferences, as well as insights for prescriptive design. While our results confirm the privacy-preserving design of most European contact tracing apps, they also provide a more nuanced understanding of acceptable features. Based on market simulation and variation analysis, we conclude that adding goal-congruent features will play an important role in fostering mass adoption.

**Introduction**

The COVID-19 pandemic has created a state of emergency in countries worldwide. Governments imposed lockdown measures to help control the outbreak of the virus and slow down its transmission. As lockdowns resulted in negative economic and social consequences, contact tracing apps arguably are one of the best tools we currently have available to avoid a second wave and potential re-lockdown. These apps have been developed in various countries, among them SwissCOVID in Switzerland, StopCOVID in France, Corona-Warn-App in Germany and many more. Experts estimate that a critical mass threshold of 60% of the country's population would be required to ensure contact tracing apps are effective (University of Oxford 2020). However, in many countries adoption rates are far from achieving this goal. For instance, Italy, the first European country with COVID-19 outbreak, has an adoption rate of 14% (Follis 2020). France's user base of StopCOVID never increased beyond an initial adoption rate of 3% (archyde 2020). Singapore with 40% adoption, as well as Germany with 21% (Cellan-Jones and Kelion 2020) and Switzerland with 17% (FOPH 2020), boast "best in class" adoption rates to date, however still fail to meet expectations in terms of critical mass. The introduction of contact tracing apps has been accompanied by controversial debates about their privacy implications. This debate is primarily conducted by experts and at the political level, but lacks consideration of the user's perspective. Among the few empirical studies that focus on users are Trang et al. (2020) who analyze the impact of various app specifications (i.e. benefit appeal, privacy design, and convenience design) on app acceptance. However, they conducted their study before the contract tracing apps were launched and therefore lack the perspective from actual users and actual designs. In view of the slow adoption rates, von Wyl et al. (2020) call for more research on the acceptability of contact tracing apps to provide an understanding of the rationale behind their use. Gupta and De Gasperis (2020) argue for the participatory design of such apps in order to improve their market acceptance. To address this gap, we study the following question:

> *What are users' preferences for contact tracing app features and how can they inform app design?*

Based on a conjoint analysis (CA) study with 300 participants in Germany, we provide empirical insights into users' preferences for core and privacy-preserving features as well as value-added services of contact tracing apps. CA, as an established market research technique, is a "practical set of methods for predicting consumer preferences for multi-attribute options in a wide variety of product and service contexts" (Green and Srinivasan 1978, p. 103). It has been occasionally used for understanding the privacy trade-offs in the design of personal ICTs (Mihale-Wilson et al. 2017; Naous and Legner 2019). Our results confirm the dominant privacy-preserving design of most national contact tracing apps in Europe, but also contribute to a more nuanced understanding of acceptable features. From our study, we gain insights for prescriptive design that allow the formation of app features that fit users' expectations, with implications for service providers to adjust their offerings to different user segments. Following market simulations, we demonstrate that goal-congruent features, and specifically value-added services, can play an important role in driving user adoption.

The remaining of the paper is structured as follows: First we provide a background on contact tracing in the context of COVID-19. Then, we introduce the applied research methodology, followed by a detailed description of the design of the CA. Afterwards, we present the empirical results. Finally, we discuss our findings and conclude with implications for research and practice.

**Background**

*Contact Tracing and Disease Control*

Contact tracing is a key control measure in the battle against infectious diseases and can break the chains of transmission when systematically applied. The World Health Organization (WHO) defines contact tracing as "the process of identifying, assessing, and managing people who have been exposed to a disease to prevent onward transmission" (WHO 2018, p. 2). Contact tracing is an extreme form of locally targeted control and has the potential to be highly effective when dealing with a low number of cases (Eames and Keeling 2003). It has been traditionally performed by health authorities using expert-led interview-based techniques.

In the case of COVID-19, contact tracing requires identifying people who may have been exposed to the virus and following up with them daily for a period of at least 14 days from the last point of exposure (Legendre et al. 2020). The fact that symptom onset may only occur days after infection makes it difficult for traditional approaches to map the network of potential exposure traces and thus control the transmission rate of the virus. Therefore, advanced techniques are required for effective contact tracing in the COVID-19 context.

*Contact Tracing Apps for COVID-19*

Governments and health authorities over the world therefore promote mobile applications that enable digital contact tracing, as fast and reliable solution to support traditional approaches performed by the public health authorities in fighting pandemics. The goal of these contact tracing apps is to continuously track user's proximity and to notify them in the event of possible COVID-19 exposure for self-isolation (Legendre et al. 2020). Simulations confirm that if approximately 60% of the population uses the requisite country app, it has the potential to stop the epidemic and keep countries out of lockdown (University of Oxford 2020).

Among the first countries to develop and launch a contact tracing app was Singapore with TraceTogether. The app has to date 2.3 million users indicating around 40% adoption rate (tracetogether.gov.sg). Based on the same framework, the Australian app (COVIDSafe) currently boasts a user base of around 7 million, which represents over a quarter of the Australian population (Norman 2020). In Europe, Austria's Stopp Corona App was first launched in March. Currently uptake is 8% remaining well below expectations (Reuters 2020). Italy, which was among the mostly affected countries with COVID-19, launched Immuni app in June, but its adoption rate remains at 14% (Follis 2020). France also launched StopCOVID in the same period, and has only 3% adoption rate (archyde 2020). Among the countries that witnessed a higher rate of adoption in Europe are Germany and Switzerland. Germany's Corona-Warn-app was launched in June and has over 17 million users (over 20% of the population) 4 months after the launch (Cellan-Jones and Kelion 2020). Switzerland also launched its SwissCOVID app in June, 3 months later, it has over 1.5 million users, however lags behind the active user goal of 3 million for SwissCOVID to be effective (FOPH 2020).

### *Design of Contact Tracing Apps for COVID-19*

The design of national contact tracing apps (Table 1) has been subject to a lively debate in most European countries with emphasis on their privacy implications. Common tracing mechanisms rely on smartphone's absolute location (in the case of location-based tracing) or relative location (in the case of proximity-based tracing) to other smartphones (Legendre et al. 2020). Proximity-based contact tracing relies on Bluetooth Low Energy (BLE) to infer the relative proximity of smartphones (up to 50m outdoors and 25m indoors), while location-based contact tracing uses GPS traces for precise location. Whereas most countries use BLE technology in building their contact tracing apps, only few have adopted a location tracing mechanism for cross-checking paths, among them the Israeli app Hamagen.

The alerting mechanism in these apps (i.e., centralized versus decentralized) has significant privacy implications (Criddle and Kelion 2020). While both approaches require a central server for exchanging the pseudo IDs of users, the matching of traces with positive user IDs is the main difference. With the centralized approach, IDs are shared with the central server managed by the public health authorities for matching with positive cases and notifications. This allows authorities to have a controlled environment for fighting the pandemic since the alerting is carried out by the central server in case of a match. With a decentralized approach, the matching is done on the user's smartphone with the list of infected IDs. Both approaches communicate anonymously, however the decentralized approach is regarded as more privacy-preserving since no logging data is exchanged with the server from the user's smartphone, except in the case of infection (Legendre et al. 2020). Among the countries that follow a centralized approach is France with the StopCOVID app, which is built based on the ROBust and privacy-presERving proximity Tracing protocol (ROBERT). It is worth noting that apps with centralized architecture might require pre–registration with personal information (e.g., TraceTogether and COVIDSafe) for verification by the central server, however, apps relying on the ROBERT protocol do not require such information (Ahmed et al. 2020).

The core functionality of apps comprises of tracing and alerting users. In addition, these apps can provide additional features for fighting the pandemic and applying safety measures. For instance, logged data on encounters and information provided on the app can be used to estimate possible infection risk, which is the case for the Corona-Warn-App that is accompanied by a risk-assessment feature. Other apps provide notifications on safe places and infected zones, or contextual services such as check-in services for safe entry (e.g., TraceTogether).

| App (by country) | Launch Date | Number of users | % of total of population | Approach | Technology | User Identification | Special Feature |
|---|---|---|---|---|---|---|---|
| 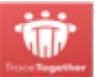 **TraceTogether (Singapore)** | March 20th | +2M | ~42% | Centralized | based on legacy BLE | Phone number required | SafeEntry integration |
| 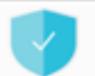 **Hamagen (Israel)** | March 22nd | +1.5M | ~17% | Decentralized | Cross-referencing of GPS data | No information required | Safe places |
| 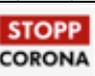 **StoppCorona (Austria)** | March 25th | +0.7M | ~8% | Decentralized | based on legacy BLE | Phone number required | Symptoms check |
| 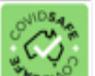 **COVIDSafe (Australia)** | April 26th | +7M | ~28% | Centralized | based on legacy BLE | Personal information required | - |
| 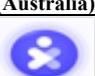 **Immuni (Italy)** | June 1st | +4M | ~14% | Decentralized | Apple-Google Exposure Notification | Region required | - |
| 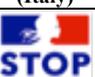 **StopCOVID (France)** | June 2nd | +2.3M | ~3% | Centralized | ROBERT (centralized based on legacy BLE) | No information required | - |
| 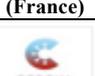 **Corona-Warn-App (Germany)** | June 16th | +17.8M | ~21% | Decentralized | Apple-Google Exposure Notification | No information required | Risk assessment |
| 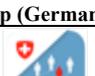 **SwissCOVID (Switzerland)** | June 25th | +1.5M active users | ~17% | Decentralized | DP-3T and Apple-Google Exposure Notification | No information required | - |
| 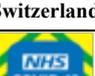 **NHS COVID-19 (UK)** | Not launched | - | - | Decentralized | Apple-Google Exposure Notification | Region required | Symptoms check |

Table 1. Overview of Contact Tracing Apps

*Research Gap*

Augmenting adoption of contact tracing apps has proven to be a challenge, and involving the users in the discussion on app characteristics and aspects related to the data processing is critical for ensuring mass acceptance. This was also highlighted by Gupta and De Gasperis (2020), who suggest participatory design with users to better build contact tracing apps. The few studies that have investigate the user perspective highlight the important role of perceived benefits for self and society together, privacy perceptions and usability aspects in the adoption of contact tracing apps (e.g., Trang et al. 2020). While those studies focus on user perceptions, they do not provide input on the most appealing realization options that inform app design. Moreover, most studies were conducted before these apps were launched; they therefore lack the perspective from actual users and actual designs. This has resulted in calls for research, for instance by von Wyl et al. (2020), on the acceptability of contact tracing apps and adherence by the target population. While Trang et al. (2020) suggest that there should be one app that fits all with defined specifications in this specific context of contact tracing, we foresee an opportunity to address the varying preferences of the whole population. Based on prior research (Wortmann et al. 2019), we can infer that goal-congruent feature additions could have a positive influence on app adoption by the foreseen benefits associated to using the app through extended services.

**Research Approach and Design**

Our study aims at improving the understanding of users' preferences and privacy trade-offs in contact tracing apps and gain additional insights for prescriptive design (Bélanger and Crossler 2011). We employ CA, which aims to provide evidence on the most influencing factors on the consumer's choice of a product. Applying the utility concept from economics, CA enables the estimation of a user preference structure based on his evaluation of different product attributes or features. It is a very suitable method to inform IS design through an empirical analysis of user preferences. For these reasons, CA is gaining popularity to study information privacy trade-offs in different types of services (Krasnova et al. 2009; Ho et al. 2010).

In applying CA, we follow the methodological guidelines for IS studies outlined by Naous and Legner (2017) and use ACBCA, which extends the traditional full-profile CA (Green and Srinivasan 1978). This CA variant combines the advantages of adaptive- and choice-based procedures (Johnson et al. 2003). It is based on a choice experiment where participants have to choose among a set of profiles (corresponding to different product combinations) after they perform a self-explicated task to assess must have and unacceptable attribute levels from the evaluation to reduce the choice burden. Based on users' choices, part-worth utilities and relative importance measures are calculated using the Hierarchical Bayes (HB) estimation (Howell 2009).

*Attributes and Levels Selection*

A first and often challenging step in CA is to identify the attributes that are relevant to users in forming their preferences. In selecting the attributes and levels, we followed a mixed method approach (Naous and Legner 2017) based on four stages.

In the first stage, we reviewed recent articles that compare the different contact tracing apps (Legendre et al. 2020; Ahmed et al. 2020) and assess the user's perspective (Trang et al. 2020; Gupta and De Gasperis 2020) to identify attributes describing the core functionalities and privacy related characteristics. This resulted in 12 attributes corresponding to four dimensions representing the main contact tracing app features: *initiation*, *core functionalities*, *transparency and control*, and *platform characteristics*. In the second phase, we examined 9 contact tracing apps (cf. Table 1) to understand the realization options and identify the attribute levels. Based on this analysis, we decided to add two attributes to our list characterized as *value-added services* that can provide additional benefits and attract more users; diagnosis and contextual services.

In the third phase of the attributes and levels selection, we organized a focus group with five current and potential users of COVID-19 apps to identify important attributes and eliminate unacceptable option. This phase also allowed us to add one attribute that we did not consider in our initial list. The ability to provide a risk assessment on the app might require additional health information for accurate estimations. We therefore consider health information registration as an option for the initiation dimension.

We finally assess the list of attributes and levels with two privacy experts (who are also familiar with the different contact tracing apps) that validated the attributes and the levels. Based on these phases, our final list was formed of ten attributes with their corresponding levels (Table 4):

The *Initiation* dimension specifies attributes related to the registration to the app. We excluded attributes for registration options that are present in all apps and focus only on one attribute:

- Health information registration: specifies whether data about health status (e.g., COVID-19 risk groups) is required on the app or not for a more robust data analysis and ideally risk assessment.

The *Core Functionalities* comprises of three basic features offered by all existing COVID-19 apps:

- Exposure logging: corresponds to the tracing mechanism employed on the app. It could be proximity tracing with Bluetooth technology, location tracking via GPS traces or a combination of both.
- Test results sharing: indicates how the exposure notification is triggered on the app; it could be via user sharing on the app symptoms or positive test results, sharing positive test results validated by the healthcare provider, or direct sharing of test results by the healthcare provider (i.e., also includes clearing status in case of negative test result).

- Exposure notification: refers to how users get notifications in case of encounter with an infected person. It could be alerting only in case of exposure, in addition users can get risk assessment based on logged data, information on country region, health status and possibly other background information.

*Value-added services* comprise features that provide additional benefits to users.

- Diagnosis services: can be used for checking COVID-19 symptoms; they can be either through basic health checklists on possible symptoms, or advanced diagnosis with machine learning on mobile sensor data (i.e., heart rate, breathing, coughing strength, etc.) (CORDIS 2020).
- Contextual services: correspond to additional services related to safety measures; examples are check-in services for safe entry in public places based on customer count or identification of safe places and infected zones through interactive maps.

*Transparency & Control* comprise features for transparent data management on the app.

- Dashboard: corresponds to the transparency about the data usage on app; could be a basic dashboard on status and data logs or detailed with sharing information on data logging, contact traces and sharing parties.
- Data sharing purpose: refers to what is the target of data sharing and with whom it will be shared; it can be restricted to contact tracing (sharing with app provider, i.e., public health authorities only), involves epidemiological insights and research (sharing with public health authorities, healthcare providers and researchers), or also includes sharing for additional safety measures (for instance check-in at restaurants, public transports or workplaces).

*Platform characteristics* correspond to the technical design of the app and the communication between the app and the remote server.

- App Architecture: corresponds to the communication between the app and the central server, which can be implemented in a centralized or decentralized approach (Ahmed et al. 2020). In a centralized architecture, users share their IDs with a central server and matching with positive cases is done on the server. In a decentralized approach, only an infected person is required to share data with the server and all matching with positive cases is done on the user's smartphone that periodically receives the list of infected IDs from the server.
- Interoperability: corresponds to the cross-country integration options; it could be a national app that can only be used in the specific country, or a national app that allows safe information exchange with other apps to be used when travelling.

| **Dimension** | **Attribute** | **Attribute Levels** |
|---|---|---|
| Initiation | Health Information registration | No information is required |
| | | Health status (i.e., COVID-19 risk groups information) |
| Core Functionalities | Exposure Logging | Contacts (Bluetooth) |
| | | Locations (GPS Traces) |
| | | Contacts & Locations |
| | Test Results Sharing | User can share symptoms or positive test results on app |
| | | User can share positive test results on app only with a validation code by the healthcare provider |
| | | Healthcare provider directly shares test results (positive/ negative) with users |
| | Exposure Notification | Alert only if you had contact with an infected person |
| | | Alert if you had contact with an infected person; includes risk assessment (low, medium, high) |
| Value-added Services | Diagnosis services | No in-app diagnosis |
| | | Simple diagnosis: Symptoms tracking with a checklist |
| | | Advanced diagnosis: Using sensors to capture symptoms (e.g., breathing, coughing) |
| | Contextual services | No additional services |
| | | Check-in service with QR code in public places for safe entry (e.g., restaurants, supermarkets) |
| | | Maps with indication of safe areas/ infected zones |

| Transparency and Control | Dashboard | Basic dashboard on data logging |
| --- | --- | --- |
| | | Detailed dashboard on data logging, updates and sharing |
| | Data sharing | Restricted to contact tracing (sharing with app provider, i.e., public health authorities) |
| | | Contact tracing, epidemiological insights and research (sharing with public health authorities, healthcare providers and researchers) |
| | | Contact tracing, research and specific purposes for safety measures (e.g., restaurants, transportation providers, workplace) |
| Platform Characteristics | App Architecture | Centralized |
| | | Decentralized |
| | Interoperability | Cross-country integration |
| | | No cross-country integration |

Table 4. List of attributes and levels

*Study Setup*

To run our study, we used Sawtooth Software, which is a specialized software with advanced modules for CA survey administration and data analysis. The online survey started with an introduction about contact tracing apps and the conjoint survey sections. We then explained the attributes involved and the different levels (or options) before collecting user choices in the typical ACBCA sections. We added screenshots of the app when possible to illustrate the differences between levels, this was done for two attributes: exposure notification and dashboard (Figure 2). Visuals would make it easier for the users to select based on concrete realization instead of verbal descriptions (Naous and Legner 2017).

Participants had to complete the four ACBCA sections in the following order:

1. **Build Your Own (BYO):** Participants are asked to build the most preferred configurations of the contact tracing app from the list of available attributes and levels. This provides input of individual preferences. The following sections are then adapted to the preferred levels selected by the participants.

2. **Screening:** The survey contained 7 screening tasks with 3 options, where participants assess the possibility of using different app designs. As part of the self-explicated task, this section helps in understanding the user's non-compensatory behaviour. Respondents are asked about must have and unacceptable features based on their response pattern. These identified features will not be displayed later to avoid bias in selection.

3. **Choice Task Tournament:** Based on their answers to previous questions, "[...] respondents are evaluating concepts that are close to their in the build your own section specified product, that they consider 'possibilities', and that strictly conform to any cut off (must have/unacceptable) rules" (SawtoothSoftware 2014). We present a maximum of 10 choice tasks to respondents with three options. This allows us to estimate the user preferences for the different attributes and levels based on the choice data.

4. **Calibration:** While traditional CBCA includes a "None" option, this is not available in ACBCA. Instead a "None" threshold can be estimated via the Screening and Calibration section. To calibrate utilities, participants are shown six concepts, including the concept identified in the BYO section, the concept winning the Choice Tournament as well as four previously shown concepts that were either accepted or rejected. The participant is asked about their likelihood to use these concepts using five-point scale from "Definitely would not" to "Definitely would".

The last phase of the survey included questions on demographics (gender, age) and professional background, as well as questions on general mobile app use and opinion about the COVID-19 app.

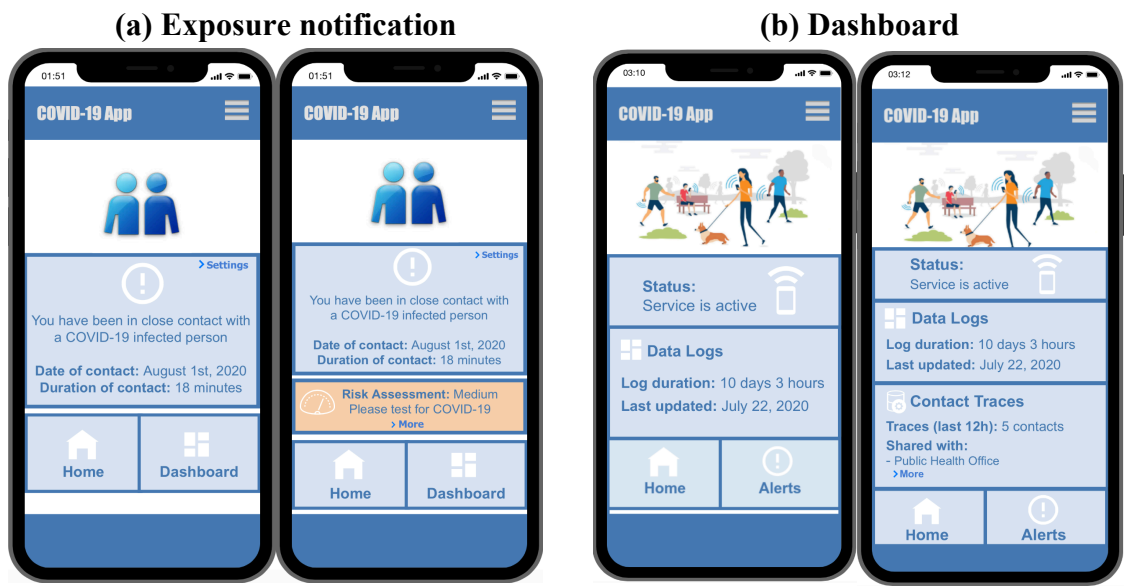

Figure 2. Mobile screenshots for attributes levels

*Study Sample*

To obtain qualified results, we targeted 300 participants from Germany, the country with the highest number of absolute contact tracing app users (now 17.8 million), who are users or potential users of the national contact tracing app (Corona-Warn-App). We selected Prolific.co as crowdsourcing platform to hire survey participants from an online pool of users. Crowdsourcing platforms, such as MTurk and Prolific, provide fast, inexpensive and convenient sampling method and are appropriate for generalizing studies (Jia et al. 2017). They have been widely used in research on security and privacy (Redmiles et al. 2019), and allow a wide reach in CA studies (Pu and Grossklags 2015; Naous and Legner 2019). The study setup was examined by the Ethics Committee within our academic context, to guarantee that respondents' participation was completely anonymous and all data collected is treated confidentially and will not be disclosed in its original form.

Participants were screened based on their smartphone use and knowledge about the COVID-19 app. Survey respondents were compensated 2.50£ for their participation, which is a fair amount for a 15-20 minutes survey on this platform. As quality criteria, we eliminated 17 responses that took less than 7 minutes for survey completion, which might affect the consistency of the analysis.

From the total remaining 283 respondents, that we included in the final data analysis, we had 55.83% male participants and 44.17% females. The majority aged between 26 and 35 years old (50.18%) and 94% less than 46 years old. Our respondents are mostly privacy aware and have frequently heard about the misuse of user information on media and press (82.33%). In terms of mobile app use, our sample is tech-savvy and uses plenty of apps, among them navigation (95.41%), social networking (79.86%) as well as health and fitness apps (54.77%). Finally, we note that 62.54% of the respondents think that the COVID-19 contact tracing app should be mandatory.

| **Variable** | **Level** | **%** |
| --- | --- | --- |
| Gender | Male | 55.83 |
| | Female | 44.17 |
| Age | 18-25 | 31.10 |
| | 26-35 | 50.18 |
| | 36-45 | 12.72 |
| | 46-55 | 3.53 |
| | 56-65 | 2.12 |
| | 66-75 | 0.35 |
| Privacy Awareness | Not informed | 17.67 |
| | Well informed | 82.33 |

Table 5. Sample demographics and background information

**Results**

*Relative Importance*

CA provides relative importance scores based on part-worth utilities for each attribute (Figure 4). Our results show exposure logging (19%) and test results sharing (13%) as the two most important attributes. The app architecture (12%) comes next, which reflects the general debate about central and decentralized architectures. Diagnosis services had 11%, as value-added service, whereas interoperability (i.e, cross-country integration) and contextual services had a similar importance of 10%. Data sharing and health information registration follow with 8% importance score, even though these two attributes are concerned about user privacy on the app and the associated risks. Although being a core service, exposure notification (5%) was less important to users who might not be interested on the method or form of the notification. Interestingly, the transparency on the app was least important with a score of 4%, which contradicts other studies on privacy concerns and transparency in data management (Ahmed et al. 2020).

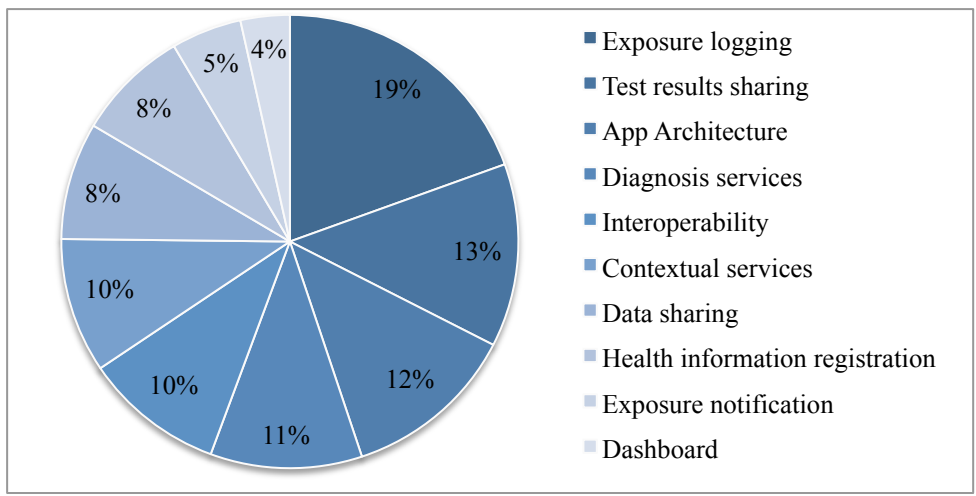

Figure 4. Relative importance of contact tracing app attributes

*Part-worth Utilities and Preferences*

Part-worth utilities are normalized HB estimates that provide insights about users' preferences for the different attributes and levels. Positive utilities correspond to preferred levels and negative utilities correspond to undesired levels. We assess the "goodness of fit" using percentage certainty (PC) and root likelihood (RLH) (Giessmann and Stanoevska 2012). We obtained a PC mean of 0.486, indicating acceptable results of fit. RLH valued 0.654, which is considered more fit than the chance level given we have three choice tasks.

The part-worth utility distribution (Table 6) allows us to identify attribute levels that are mostly selected by users through the choice options, thus correspond to their preference structure and trade-offs with respect to the overall app design. Interestingly, we observe that users prefer to provide information about their health status on the app, most likely because this information would help to provide more targeted analysis of their situation in regards to COVID-19. In terms of exposure logging, contact tracing via Bluetooth - the most privacy-preserving option - has the highest utility, while GPS tracking had a negative utility and a combination of both has positive utility. For test results sharing, users have positive utilities for trusted and officially validated test results sharing. However, the highest utility was for sharing by the user via a validated code from the healthcare provider. For exposure notification, users appreciate having a risk assessment in addition to the notification. In terms of value-added services, the highest utilities were for simple diagnosis service. Although advanced diagnosis options with mobile sensors can be of great help in detecting patterns and assessing severity of symptoms, users seem to have concerns of extensive data collection via the app. For contextual services, users prefer the second option with maps identifying infected zones. However, when assessed individually in the BYO section, users stated that they would not prefer an additional contextual service with the app. For transparency and control, higher utilities were recorded for the detailed dashboard and restricted data sharing, which are more privacy-preserving options. For the choice of platform, users have positive utilities for the decentralized approach as more privacy-preserving approach also. Finally, the cross-country integration is preferred by users. Our results thereby support the European Union (EU) member states' effort to establish a technical framework for cross-country contacts tracing for travelers and cross-border employees (Lomas 2020).

| Attribute | Attribute levels | Average Utilities | Standard Deviation | Distribution for BYO Section (%) |
|---|---|---|---|---|
| **Health information registration** | No information is required | -2.86 | 51.16 | 43.46 |
| | Health status | 2.86 | 51.16 | 56.54 |
| **Exposure logging** | Contacts | 41.46 | 113.56 | 46.64 |
| | Location | -50.00 | 83.95 | 15.90 |
| | Contacts and Location | 8.54 | 62.07 | 37.46 |
| **Test results sharing** | Symptoms or positive test results | -51.42 | 58.06 | 12.37 |
| | Positive test results with validation code | 32.69 | 42.74 | 48.06 |
| | Healthcare provider shares test results | 18.72 | 51.74 | 39.58 |
| **Exposure notification** | Contact with an infected person | -7.01 | 30.56 | 40.28 |
| | With risk assessment | 7.01 | 30.56 | 59.72 |
| **Diagnosis services** | No in-app diagnosis | 5.74 | 53.54 | 34.63 |
| | Simple diagnosis | 25.83 | 31.15 | 47.35 |
| | Advanced diagnosis | -31.57 | 57.32 | 18.02 |
| **Contextual services** | No additional services | -4.52 | 51.02 | 37.10 |
| | Check-in service with QR code | -8.67 | 52.58 | 29.33 |
| | Maps of safe areas/ infected zones | 13.19 | 32.43 | 33.57 |
| **Dashboard** | Basic dashboard | -9.30 | 18.76 | 37.81 |
| | Detailed dashboard | 9.30 | 18.76 | 62.19 |
| **Data sharing** | Restricted to contact tracing | 11.12 | 41.96 | 39.93 |
| | Includes epidemiological insights and research | 3.39 | 26.85 | 24.38 |
| | Includes specific purposes for safety measures | -14.51 | 46.59 | 35.69 |
| **App Architecture** | Centralized | -37.37 | 69.83 | 37.10 |
| | Decentralized | 37.37 | 69.83 | 62.90 |
| **Interoperability** | No cross-country integration | -45.09 | 44.76 | 14.13 |
| | Cross-country integration | 45.09 | 44.76 | 85.87 |

Table 6. User preferences and part-worth utilities (preferred levels are highlighted)

*User Segmentation*

To gain insights into user segments for contact tracing apps, we performed a cluster analysis based on the individual part-worth utilities. By applying k-means clustering, we derived three clusters of users with varying preferences with respect to privacy-preserving features and value-added services (Table 7). While the first two clusters (with majority of users combined) are privacy concerned and prefer basic features that guarantee user privacy, the third cluster is unconcerned and would prefer all options that provide an enhanced app.

|  | *Cluster 1* | *Cluster 2* | *Cluster 3* |
|---|---|---|---|
| Number of participants | 76 (26.85%) | 92 (32.51%) | 115 (40.64%) |
| Privacy Characterization | Privacy concerned users | Privacy concerned users | Unconcerned users |
| Value-added services | No additional services | Included | Included |
| *Preferences* | | | |
| **Health information registration** | Not required | Not required | Health status |
| **Exposure logging** | Contacts | Contacts | Contacts and Location |
| **Test results sharing** | Positive test results on with validation code | Positive test results on with validation code | Healthcare provider shares test results |
| **Exposure notification** | Contact with an infected person | With risk assessment | With risk assessment |
| **Diagnosis services** | No in-app diagnosis | Simple diagnosis | Advanced diagnosis |
| **Contextual services** | No additional services | Maps of safe areas/ infected zones | Maps of safe areas/ infected zones |
| **Dashboard** | Detailed dashboard | Detailed dashboard | Detailed dashboard |
| **Data sharing** | Restricted to contact tracing | Restricted to contact tracing | Specific purposes for safety measures |
| **App Architecture** | Decentralized | Decentralized | Centralized |
| **Interoperability** | Cross-country integration | Cross-country integration | Cross-country integration |

Table 7. Identified clusters with preferences based on customer segmentation

The first two clusters are similar by their preferences to privacy-preserving features when it comes to the core functionalities including contact tracing via Bluetooth and sharing only validated test results to avoid false alerts. However, for exposure notification, the second group prefers having a risk assessment in addition to the notification. The main difference is in the value-added services, where the first segment (26.85%) does not prefer any value added-service, while the second segment (32.51%) prefers at least a simple diagnosis service for tracking COVID-19 symptoms, and a contextual service that provides information about infected zones and safe places. For all other features, both segments share the same preferences: They do not prefer to share any health information on the app, prefer a detailed dashboard and no data sharing with parties other than the public health authorities. They also prefer a decentralized approach, however a cross-country integration.

The third cluster, with more than 40% of the participants, prefers enhanced features on all attributes. Major differences to the previous segments are in the health information registration, exposure logging, and diagnosis services where this segment would prefer a combination of contact and location tracking, as well as advanced diagnosis services. This segment also has inherent trust in the authorities, and would choose all available app features even if it would be privacy intrusive. This is shown in their choice of test results sharing by the authorities and the centralized approach. In addition, data sharing for this segment can be for the purposes of helping to fight the pandemic in different contexts.

*Variation Analysis*

Variation analysis allows us to study the effect of changing attributes on market share predictions. Thus, it provides a market simulation based on reliable quantitative data that can feed the design of the app and identify features that would improve the adoption.

With market simulation we can then understand if adding value-added services with the proposed contact tracing app can result in higher market shares, thus better adoption rates. As a reference app, we use the characteristics of the German Corona-Warn-App. We then propose 5 variations (Table 8) corresponding to the multiple combinations of value-added services within the app. App 1 has a simple diagnosis service for checking symptoms via checklists. App 2 has an advanced diagnosis service based on data processing of sensor data (e.g., heart rate, breathing, coughing, etc.) and applying machine learning algorithms on that. App 3 has a safe entry check-in service with QR code that can be used in public places for tracking the count of people inside a place and tracking positive check-ins. App 4 has a map function with indications of safe places and infected zones within a region. The final app (App 5) combines two value-added services that are selected with highest utilities: simple diagnosis and map function.

| Label | Reference | App 1 | App 2 | App 3 | App 4 | App 5 |
|---|---|---|---|---|---|---|
| Description | Corona-Warn-App | Simple Diagnosis | Advanced Diagnosis | Check-in Service | Maps | Simple Diagnosis + Maps |
| Health information registration | No information is required | | | | | |
| Exposure logging | Contacts (via Bluetooth) | | | | | |
| Test results sharing | User can share positive test results on app only with a validation code by the healthcare provider | | | | | |
| Exposure notification | Alert if you had contact with an infected person with risk assessment | | | | | |
| Dashboard | Basic dashboard on data logging | | | | | |
| Data sharing | Restricted to contact tracing | | | | | |
| App Architecture | De-centralized | | | | | |
| Interoperability | No cross-country integration | | | | | |
| Diagnosis services | No in-app diagnosis | Simple diagnosis: Symptoms tracking with checklists | Advanced diagnosis: Using sensors to capture symptoms | No in-app diagnosis | No in-app diagnosis | Simple diagnosis: Symptoms tracking with checklists |
| Contextual services | No additional services | No additional services | No additional services | Check-in service with QR code in public places for safe entry | Maps with indication of safe areas/ infected zones | Maps with indication of safe areas/ infected zones |
| Market share | | 57% | 39% | 49% | 56% | 60% |

Table 8. Scenarios for variation analysis simulation

Based on the simulation results, we observe that all apps generate market shares. This means that their utility is higher than the None threshold[1], and people would be willing to adopt such apps. However, the difference in market shares compared to the reference app (i.e., Corona-Warn-App) vary in strength. We observe that App 1 (simple diagnosis) and App 4 (Maps) would result in higher market shares with slightly better results for App 1. Consequently, App 5 with a diagnosis service of symptoms tracking and contextual service of maps also resulted in higher market shares corresponding to 60% of users.

---

[1] With the ACBCA, a None parameter can be estimated in market simulations to predict whether the respondents would be selecting a proposed option or not. Based on that, if the utility of the product concept proposed is higher than the None utility, it will be chosen.

**Discussion: Users' Preferences for Contact Tracing Apps**

The results from our conjoint analysis provide a micro perspective (i.e., that of the user) into users' preferences for contact tracing apps through evaluation of feasible design options. We thereby address Pillar III of Von Wyl et al.(2020)'s research agenda for digital tracing apps, and contribute to understanding the acceptability of these apps. Our empirical study with CA provides a system evaluation through features, and highlights which privacy-preserving features are required or mostly valued by users via part-worth utility measures and relative importance of features.

With regard to individual user preferences, we find that the exposure logging and test results sharing are the most important features in the contact tracing apps, while exposure notification as the third core service is lagging far behind. Our findings support the dominant privacy-preserving design of most European contact tracing apps; they clearly confirm user preferences for a decentralized approach and for contact tracing through proximity rather than a location-based tracking via GPS. Despite the ongoing debate about privacy concerns raised by contact tracing apps, the results show that not all privacy-preserving features are valued by users and that users care less about privacy-related aspects in comparison to core and value-added services. An alternative interpretation is that users trust these apps because they implement privacy by design principles and are implemented by authorities who protect the privacy rights of citizens through the EU General Data Protection Regulation (GDPR) (Yang et al. 2020).

Our study also provides interesting insights into the behaviour of heterogeneous respondent groups, represented by the user segments identified. The segmentation is particularly interesting given the diverging adoption rates and provides valuable insights into the prescriptive design of contact tracing apps. We realize that there is no one app that fits all (Trang et al. 2020), and that different specifications of tracing apps contribute to their mass acceptance. While we observe a tendency for privacy-preserving features and basic functionalities (in the first and second segments), the largest segment of users values extended services more than privacy-enhancing features. Our simulation results based on this CA approach highlight how value-added services are an important topic for consideration in further developing and improving the current apps through targeted and extended services.

*Implications*

By providing empirical insights into individual and group preferences, our results draw the attention to heterogeneous types of users with diverging preferences. We argue that these user segments need to be addressed with targeted features to achieve mass adoption, and thereby contradict Trang et al. (2020)'s view that there can only be one set of specifications of a tracing app for all citizens. Based on the market simulation and variation analysis, we conclude that contact tracing apps could achieve higher market shares if value-added services were added beyond the basic app for tracing encounters. This is in line with studies by Wortmann et al. (2019) who have shown that adding goal-congruent features to a core system may result in higher adoption. For contact tracing apps, goal-congruent design would imply a paradigm shift from a strong focus on privacy-aware design to explore value-added services that complement the app (e.g., through diagnosis and contextual services). Based on our results, value-added services could become a game changer in the adoption challenge. A viable implementation option that would take into account privacy preferences is to develop auxiliary apps that can be integrated within the COVID-19 app if needed. Similar to Singapore, who has merged the national TraceTogether app with the SafeEntry app as part of the safety measures to increase adoption rate (Lee 2020).

From a practical perspective, our results are relevant to application developers and service providers of contact tracing apps. The preference model resulting from the CA study provides concrete realization options of the contact tracing app to be taken into consideration in order to gain sufficient critical mass and acceptability amongst users. Our study supports the idea of participatory design (Gupta and De Gasperis 2020) through providing a data-driven approach that allows capturing user preferences and including the different stakeholders in the discussion of most convenient design options.

*Limitations*

Contact tracing apps have a national scope and thus may be impacted by the specific national implementation as well as contextual factors. Therefore, an important limitation of our study is its focus on a sample from Germany with apriori model of decentralized contact tracing. It would be interesting to have comparative studies in other countries that have introduced centralized proximity or location-based tracing apps to assess the different design options.